# Optimization of Enzymatic Biochemical Logic for Noise Reduction and Scalability: How Many Biocomputing Gates Can Be Interconnected in a Circuit?

**Vladimir Privman, Guinevere Strack, Dmitry Solenov,**
**Marcos Pita** and **Evgeny Katz***

*Department of Chemistry and Biomolecular Science and*
*Department of Physics, Clarkson University, Potsdam NY 13699, USA*

*Corresponding author: ekatz@clarkson.edu

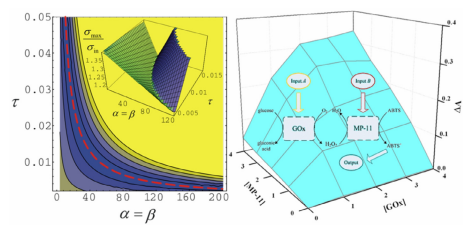

Experimental and theoretical study of an enzyme-logic AND gate suggests the possibility of a stable operation for several concatenated biocomputing gates, when properly optimized.

**Abstract**

We report an experimental evaluation of the "input-output surface" for a biochemical AND gate. The obtained data are modeled within the rate-equation approach, with the aim to map out the gate function and cast it in the language of logic variables appropriate for analysis of Boolean logic for scalability. In order to minimize "analog" noise, we consider a theoretical approach for determining an optimal set for the process parameters to minimize "analog" noise amplification for gate concatenation. We establish that under optimized conditions, presently studied biochemical gates can be concatenated for up to order 10 processing steps. Beyond that, new paradigms for avoiding noise build-up will have to be developed. We offer a general discussion of the ideas and possible future challenges for both experimental and theoretical research for advancing scalable biochemical computing.

# 1. Introduction

**1.1. Background.** There has been a significant recent interest in the emerging field of chemical computing.[1] Many different chemical systems have been developed to mimic operations of various electronic elements such as logic gates,[2] switches,[3] memory units[4] (e.g., read-write-erase[5] and flip-flop systems[6]). Chemical computing systems have been realized in solutions[7] or assembled at chemically modified interfaces.[8] A composition of chemical information processing systems varies from simple signal-responsive molecules or polymers[9] to very sophisticated supra-molecular "machines" performing complex operations upon molecular translocations of their parts.[10]

Many different external physical signals, including optical,[11] electronic,[12] magnetic[13] and/or chemical inputs (e.g., variation of pH),[14] were applied to activate chemical processes mimicking computational operations. It should be noted that most chemical computing systems produce chemical output signals (products of chemical reactions) and these signals should be used as inputs for the next chemical computing operation when the individual chemical steps are connected in a complex chemical information processing network. Assembling of chemical computing units in multifunctional systems resulted in chemical "devices" performing simple arithmetic operations: half-adder/half-subtractor,[15] full-adder/full-subtractor,[16] and mimicking electronic units, e.g., digital demultiplexer[17] and keypad lock.[18]

Further increase of complexity of chemical computing systems has resulted in the integration of single-function "devices" into complex multi-functional computing networks composed of many logic elements.[19] Miniaturization of chemical computing systems should lead to logic gates operating at a level of a single molecule[20] allowing for information processing at the nano-scale.[21] Great future potential has been envisaged for chemical computing due to rapid progress in the field.[22]

Chemical computing based on biomolecular systems (biocomputing, or biochemical computing) offers exciting prospective applications due to the biocompatibility of biological materials and their unique properties[23] as chemical reactants, especially, their specificity. Biocomputing systems can involve DNA,[24] proteins/enzymes,[25] and whole cells,[26] utilizing their biorecognition, biocatalytic and bioregulation properties. Recently developed logic gates based on enzymatic reactions[27,28] utilize the specificity of biocatalytic processes allowing several simultaneous steps in one solution without interference and "cross-talk." Using the advantages of enzyme-based logic gates, biocomputing "devices" such as a biomolecular half-adder, a half-subtractor[28] and keypad lock[29] were assembled to demonstrate systems with computational functionality composed of several concatenated logic gates.[29,30]

**1.2. Motivation.** With the advent of biochemical computing, it has become important to explore scalability of biochemical logic gates.[31] Error-generation and its reduction by noise minimization have to be explored both experimentally and theoretically. The aim of the present work is to initiate such a study, on an example of an enzyme-involving AND gate.

Biocomputing gates frequently mimic Boolean logic gates. The reason for this has been the expectation that biochemical logic will ultimately be coupled to ordinary electronics in applications. Furthermore, it is hoped that scalability paradigms for complex information processing can be adapted from ordinary electronics to biochemical logic. For definiteness, let us assume that two chemicals, of concentrations $A$ and $B$, are the input signals for a reaction that mimics a logic gate and yields the third chemical, of concentration $C$, as the output signal. We will set the Boolean 0 as zero concentrations of all the chemicals, whereas the Boolean 1 values can be selected as some experimentally convenient (or set by gate concatenation, as explained later) values $A_{\max}$, $B_{\max}$ and $C_{\max}$.

Let us now define the "logic" (dimensionless) concentrations,

$$x = A/A_{\max}, \qquad y = B/B_{\max}, \qquad z = C/C_{\max}. \tag{1}$$

Of primary interest in Boolean logic are the values of $x, y, z$ near 0 and 1. However, the biochemical reaction should actually be considered for all possible values of the concentrations, and described by a function

$$z = F(x, y). \tag{2}$$

This function depends not only on the arguments $x, y$, but also parametrically on $A_{\max}$, $B_{\max}$, as well as on the reaction time $t_{\max}$, and other biochemical-system parameters, such as various reaction rates. In practice, the function

$$F(x, y; A_{\max}, B_{\max}, t_{\max}, ...) = C(x, y; A_{\max}, B_{\max}, t_{\max}, ...) / C(1,1; A_{\max}, B_{\max}, t_{\max}, ...), \tag{3}$$

will usually not be exactly known but only phenomenologically modeled and fitted from experimental data, as illustrated in Section 3. We note that $C_{max} = C(x=1, y=1; A_{max}, B_{max}, t_{max}, ...)$, and therefore it does not constitute an adjustable parameter. Furthermore, in fitting the function $C$, the dependence on the first four arguments in equation 3 is simply via the products $xA_{max}, yB_{max}$ (which are the concentrations $A, B$). Therefore, $A_{max}, B_{max}$ are also not adjustable parameters when fitting the reaction kinetics data.

The parameters $A_{max}, B_{max}$ only enter when we consider the logic function of the gate in the optimization step. Our aim is to develop, see Section 3, approaches for optimizing the set of values $\{A_{max}, B_{max}, t_{max}, ...\}$, where the dots here and in equation 3 refer to reaction parameters, to minimize noise buildup. One can, of course, consider other optimization ideas, such as redefining not only the logic-1, but also allowing for a shift of the logic-0 from the physical zero values of the concentrations, etc. However, we limit ourselves to the above formulation for definiteness. In fact, if the gate is to be used in concatenation with other gates in a logic circuit, then the values of $A_{max}, B_{max}$ might be set by the connected processes. Thus, optimization can be carried out only with respect to those parameter combinations that can be changed by varying values from the subset $\{t_{max}, ...\}$, assuming that they can be controlled in the available experimental setting. This is illustrated in Section 3.

We point out that there can be many sources of noise in the process considered. The input signals may not be exactly at the logic values 0 and 1. The output can also have additional noise due to the nature of the reaction: technically, the function $F(x,y)$ may actually be probabilistically distributed due to a random component in it. However, the main source of possible *noise amplification* during the biochemical gate function, is simply due to the shape of the function $F(x,y)$ in the vicinity of the four logic argument values $x, y = (0,0), (0,1), (1,0), (1,1)$. Indeed, if the function $F$ has large gradient values near these argument values, then any noise and fluctuations in the inputs $A, B$ will be amplified when manifesting as noise in the output $C$.

Consider "analog" noise in the output resulting from small fluctuations in the input and controlled by the shape of the gate function. In order to prevent noise amplification when several gates are combined together, small gradient values are needed near the four "logic" input points. In Nature this is accomplished by the "sigmoid" response shape. However, for really large-scale so-called fault-tolerant device operation with numerous gates involved, another mode of noise build-up, termed "digital," will become important: No matter how narrow are the statistical distributions of the input signals centered at the "logic" values, and how accurately is the function $F$ realized experimentally, there will always be some, perhaps minute probability of a really large fluctuation in the output, that will randomly yield a wrong logic (digital) value. In very large scale networks of combined gates, such a "digital" mechanism of error build-up becomes dominant, and additional error correction techniques, based on redundancy, and not reviewed here, have to be utilized.[32]

Presently biochemical computing is at least a couple of orders of magnitude away from the number of combined gates that would necessitate "digital" error correction. Thus, in the present paper we exemplify analysis of experimental data obtained for a biochemical gate, for minimizing "analog" error amplification. The specific biochemical AND gate considered, was carried out by utilizing two biocatalysts as inputs: glucose oxidase (GOx) and microperoxidase (MP-11). They catalyzed the following reactions, see Figure 1,

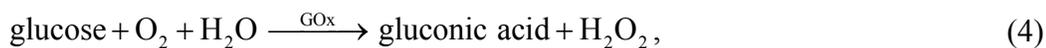

$$\text{glucose} + O_2 + H_2O \xrightarrow{\text{GOx}} \text{gluconic acid} + H_2O_2, \qquad (4)$$

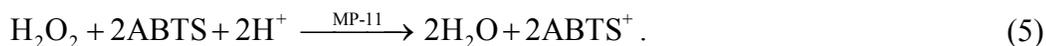

$$H_2O_2 + 2\text{ABTS} + 2H^+ \xrightarrow{\text{MP-11}} 2H_2O + 2\text{ABTS}^+. \qquad (5)$$

The actual structures of the compounds involved are shown in Figure 1. The output signal was measured as the concentration of the oxidized dye, $\text{ABTS}^+$, by optical means and defined as the change of the absorbance, $\Delta A$, at a specific wavelength.

The outline of the rest of the paper is as follows. In Section 2, we summarize the experimental procedure and results for the biochemical AND gate considered. In Section 3, we introduce methods to analyze a gate-function for error amplification, and suggest possible approaches for gate optimization. In particular, in Section 3 the experimental data are fitted to yield a function of the type of equation 2. Section 4 offers conclusions and avenues for future work.

## 2. Experimental

**2.1. Chemicals and reagents.** The biocatalysts and other chemicals were purchased from Sigma-Aldrich and used as supplied: glucose oxidase (GOx) from *Aspergillus niger* type X-S (E.C. 1.1.3.4), microperoxidase-11 (MP-11) prepared by enzymatic degradation of equine heart cytochrome c, β-D-(+)-glucose (99.5% GC), 2,2'-azino-*bis*(3-ethylbenzthiazoline-6-sulphonic acid) (ABTS). Ultrapure water from NANOpure Diamond (Barnstead) source was used in all of the experiments.

**2.2. The chemical logic gate and input signals.** The AND gate, see equations 4 and 5, was carried out in an aqueous solution consisting of glucose, 0.1 M, oxygen (initially in equilibrium with air), and ABTS, 0.1 mM, in 0.01 M phosphate buffer, pH 5.0, total volume 1 mL. Soluble GOx and/or MP-11 were used as biocatalytic input signals to activate the AND gate. The absence and presence of each biocatalyst was considered as input signals 0 and 1, respectively. The concentrations of the biocatalysts operating as input signals were varied: [GOx] = n×3.1×10$^{-9}$ M and [MP-11] = n×2.7×10$^{-7}$ M, where n = 0, 1, 2, 3 or 4, in order to map the function $F$, see equation 2. This yielded 25 data points. When one or both of the inputs was equal to 0 (9 experimental points), there was no significant change in absorbance at 415 nm. This was considered as an output equal to logic-0. However when both biocatalysts were added to the solution (16 experimental points), a change in the absorbance was observed and considered as an output equal to logic-1.

**2.3. Measurements.** The reaction took place in a 1 mL polyacrylamide cuvette in 0.01 M phosphate buffer, pH = 5.0, upon addition of biocatalyst inputs in different combinations, and we measured ΔA at λ = 415 nm after time $t_{max}$ = 145 s as the output signal. A reference cuvette was filled with the same composition as the test cuvette prior to the addition of the enzyme inputs, thus allowing the differential spectra measurements corresponding to the changes in the output composition originating from the biocatalyzed reactions. The absorbance measurements were performed using a UV-2401PC/2501PC UV-visible spectrophotometer (Shimadzu, Tokyo, Japan) at 25±2 °C.

**2.4. Experimental results.** The chemical logic gate mimicking Boolean AND was initiated in a solution containing glucose, oxygen and ABTS. Biocatalytic oxidation of ABTS resulted in the increased absorbance of the solution in the range of λ =

400-440 nm: ΔA was measured at $\lambda$ = 415 nm.[33] ABTS is oxidized only in the presence of MP-11 as the biocatalytic input and $H_2O_2$ as the oxidizer. The latter was produced *in situ* by another biocatalytic reaction due to GOx. Figure 1 shows a general scheme of the biocatalytic reactions involved in the operation of this logic gate. In the absence of the biocatalysts (input signal 0,0) or upon addition of any of the two of them, GOx or MP-11, *separately* (input signals 0,1 or 1,0) the system does not oxidize ABTS, and the solution does not yield an appreciable absorbance change, see Figure 2(A), curves a,b,c. It should be noted that the addition of MP-11 as an input signal yields an absorbance band at $\lambda_{max}$ = 397 nm as the result of the absorbance of the heme group in MP-11,[34] see Figure 2(A), curve b. This introduces a minor interference with the absorbance generated upon the biocatalytic oxidation of ABTS.

Only the addition of both biocatalysts, GOx and MP-11 (input signal 1,1), results in the formation of $H_2O_2$ and then in the oxidation of ABTS yielding the large absorbance change, Figure 2(A), curve d. The observed output signals, ΔA, correspond to the "truth table" of the AND logic gate, see Figure 2(B). Note that the data in Figure 2 corresponds to the input "intensity" n = 4 (the logic-1, as defined in Subsection 2.2) for both biocatalysts. The full set of our experimental data is given in Figure 3. These data were used to map out the function defined in equation 2, as detailed in Section 3.

**3. Reaction Kinetics and Gate Optimization**

**3.1. Reaction kinetics.** In order to map out the function $F$ defined in equation 2, we first model the reaction kinetics of the processes summarized in equations 4, 5. Our aim is to use a phenomenological rate-equation description that involves a relatively small number of (one or two) adjustable parameters. The reason for the latter is that typical data that map[35] a biochemical "input-output surface", such as shown in Figure 3, are not accurate enough for a multi-parameter fit. In fact, the strategy in Systems Biology has been similarly pragmatic: the use of phenomenological few-variable Hill functions.[36]

Let us start with the rate equation for the concentration of $O_2$, to be denoted by $O(t)$, where $t$ is the time variable. The reaction container was sufficiently small to neglect spatial variation of concentrations of the components due to external factors. However, the supply of oxygen by transport through the surface is relatively slow and constitutes a rate-limiting process. With $\bar{O} = 2.7 \times 10^{-4}$ M denoting the equilibrium value for room temperature,[37] we obtain

$$\frac{dO}{dt} = R_1(\bar{O} - O) - R_2 AO, \qquad \text{with } O(t=0) = \bar{O}. \tag{6}$$

where $R_1 = \Omega S/V$, $S$ and $V$ are the surface area and volume of the solution, respectively, equal to 61.1 mm$^2$ and 1.0 mL for the experiment described above. The quantity $\Omega$ stands for the oxygen mass transfer coefficient in the surface layer, $\Omega = 1.3 \times 10^{-3}$ mm/s, where among the published values[38,39] we took the one[39] measured under the conditions of primarily diffusional, without the added convection-mediated, transport. Note that, via its geometry dependence, $R_1$ could be used as one of the tunable parameters for gate function optimization.

Recall that $A$ = [GOx], and for the purpose of calculating the time-dependent kinetics, this concentration is a constant. Since the concentration of glucose in our case was always in great access for logic-1, we did not use a Michaelis-Menten type description[40] of the biocatalytic action of GOx, equation 4. Furthermore, the concentration of glucose did not change appreciably during the reaction. Therefore, we use the simple form shown for the second, oxygen-consumption term in equation 6. We point out that $R_2$ will be a constant to be fitted from the data. Furthermore, in the regime of applicability of the present rate-equation description, this parameter is also controllable to some degree for gate function optimization. In particular, it will depend on the concentration of glucose.

Let us now consider the concentration of hydrogen peroxide, to be denoted by $W(t) = [H_2O_2]$, with $W(0) = 0$. We also recall that $C(t)$ = [ABTS$^+$], $C(0) = 0$, and $B$ = [MP-11]. We will use the rate equations

$$\frac{dW}{dt} = R_2 AO - R_3 BW, \tag{7}$$

$$\frac{dC}{dt} = 2R_3 BW. \tag{8}$$

Here again, in order to keep the number of fit parameters down to two, we did not use a Michaelis-Menten type description of the biocatalytic action of MP-11, equation 5. This entails an assumption of access of $H_2O_2$ (when it is produced by the first reaction of the gate function), which seems to hold in the present range of experimental conditions, as confirmed a posteriori by the fact that our model fits the data. Furthermore, the amount of the dye (ABTS) oxidized in the reaction is only a small fraction of the initial concentration. Therefore, we ignore the variation of [ABTS] with time. Thus, $R_3$ will be the second constant to be fitted from the data, and it can also be tuned for gate optimization, similarly to $R_2$. Specifically, $R_3$ will depend on the concentration of ABTS.

**3.2. The gate function.** It is convenient to work with dimensionless combinations of variables,

$$a \equiv R_2 A / R_1, \qquad b \equiv R_3 B / R_1, \qquad \varsigma \equiv R_1 t, \qquad (9)$$

$$o(\varsigma) \equiv O(t)/\bar{O}, \qquad w(\varsigma) \equiv W(t)/\bar{O}, \qquad c(\varsigma) \equiv C(t)/\bar{O}. \qquad (10)$$

The system of differential equations to solve is then

$$do/d\varsigma = (1-o) - ao, \qquad \text{with } o(\varsigma=0)=1, \qquad (11)$$

$$dw/d\varsigma = ao - bw, \qquad \text{with } w(\varsigma=0)=0, \qquad (12)$$

$$dc/d\varsigma = 2bw, \qquad \text{with } c(\varsigma=0)=0. \qquad (13)$$

The solution is

$$c(a,b,\varsigma) = -\frac{2a}{1+a}\left(\frac{a}{b-a-1}+\frac{1}{b}\right)(1-e^{-b\varsigma}) + \frac{2b}{b-a-1}\left(\frac{a}{1+a}\right)^2 (1-e^{-(1+a)\varsigma}) + \frac{2a\varsigma}{1+a}. \qquad (14)$$

Note that this expression is simply a result for the function $C(A,B;\bar{O},t,R_1,R_2,R_3)$, written in a dimensionless notation. The first pair of arguments are the actual variables, which were experimentally varied over a set of 25 pairs of values, as detailed in Section 2. The parametric dependence involves the known values of $t$, which will be set to $t_{\max}$, and $\bar{O}$, $R_1$, which are both known from the literature, as well as the two unknown parameters $R_2$, $R_3$. The last datum of information needed was the conversion factor from the measured deviation in the absorbance, $\Delta A$, and the concentration of the oxidized dye, $C$. We used $C = \Delta A \times 2.7 \times 10^{-5}$ M, taking into account the extinction coefficient of $ABTS^+$ at 415 nm and the optical path length of 1 cm in our experiment.[33]

With all this information collected, we then used a least-squares fit to obtain estimates for the two unknown parameters, $R_2 = 4.7 \times 10^6 \, M^{-1} s^{-1}$, $R_3 = 1.3 \times 10^2 \, M^{-1} s^{-1}$. The estimated accuracy is approximately 10% in both rate constants. The resulting "input-output surface" is shown in Figure 4(A), in terms of $\Delta A$, with the experimental data points included. Locally, especially for small $\Delta A$, the quality of the fit for each individual point is not especially high, but for gate optimization we need a semi-quantitatively valid fit that captures the expected features of the function $F(x, y; ...)$, ultimately in various limits and over a relevant range of its arguments and tunable parameter values.

Finally, for the gate function, equation 2, we get the expression

$$F(x, y) = \frac{c(\alpha x, \beta y, \tau)}{c(\alpha, \beta, \tau)}, \quad (15)$$

where the dependence on several parameters was conveniently lumped into three dimensionless variables,

$$\alpha = R_2 A_{max} / R_1, \qquad \beta = R_3 B_{max} / R_1, \qquad \tau = R_1 t_{max}. \quad (16)$$

In fact, our experimental conditions and results of the data fit correspond to $\alpha = 746$, $\beta = 1.79$, and $\tau = 0.0115$. The properties of the gate function with these parameter values are shown in Figure 4(B). All three parameters, $\alpha, \beta, \tau$ can be modified if required for gate function optimization without affecting the values of $A_{max}, B_{max}$.

**3.3. Optimization of the logic gate.** The fitted gate function shown in the inset in Figure 4(B) significantly amplifies errors. This is specifically relevant at the logic-input $x = 0$, $y = 1$. Indeed, plots of the absolute values of the gradients of the function along various lines, in Figure 4(B), suggest that small deviations away from the line $x = 0$ in the *xy*-plane will be amplified by a factor of up to approximately 4.84. Thus, parameter optimization could involve reduction of the gradient near logic-01, perhaps at the expense of allowing for larger gradients at other logic-points.

The inputs *x* and *y* during the gate function fall near a logic point. Nevertheless, they are not set precisely: rather, there will be some probability distributions, $X(x)$ and $Y(y)$ near the logic values, $x = 0$ or $1$, and $y = 0$ or $1$. As a result, the output density will also be distributed, according to the distribution function $Z(z)$, given by

$$Z(z)dz = \int_{dS} dxdy X(x)Y(y). \quad (17)$$

Here the integration over $dS$ covers the area in the $xy$ plane that corresponds to the values of $z < F(x,y) < z + dz$, as illustrated in Figure 5. Technically, the distribution $Z(z)$ will depend on which of the four logic-pairs of $x, y$ values was inputted.

From equation 17, we get the moments expression,

$$\langle z^m \rangle = \int dxdy [F(x,y)]^m X(x)Y(y). \quad (18)$$

Here $\langle ... \rangle$ denotes averaging with respect to $Z(z)$. To estimate the noise propagation, we investigate the width of the output distribution, $\sigma_{out}$, as compared to the width of the input distributions, $\sigma_{in}$, where the latter were, for simplicity, assumed the same for both $x$ and $y$ inputs. Ordinarily in studies of noise effects one assumes that all distributions are approximately Gaussian: we took the (half-)Gaussian shape of the input, $X(x) = 2\exp\left(-x^2/2\sigma_{in}^2\right)/\sqrt{2\pi\sigma_{in}^2}$, at logic-0, and $X(x) = \exp\left(-(x-1)^2/2\sigma_{in}^2\right)/\sqrt{2\pi\sigma_{in}^2}$ at logic-1, with the same expressions for $Y(y)$. Therefore, for approximately symmetric distributions one can use the standard deviation (dispersion) definition,

$$\sigma_{out} = \langle z^2 \rangle - \langle z \rangle^2 \quad \text{(for logic-1)}, \quad (19)$$

and similarly for $\sigma_{in}$. However, for values of $x, y, z$ near the logic-0, the distributions are not centered. They are approximately half-Gaussian because negative values are not possible. For these logical points we use instead

$$\sigma_{out} = \langle z^2 \rangle \quad \text{(for logic-0)}, \quad (20)$$

and similarly for $\sigma_{in}$.

Let us define

$$\sigma_{max} = \max_{00,01,10,11} \{\sigma_{out}\}. \quad (21)$$

Ideally, to optimize a gate, we should calculate the ratios $\sigma_{out}/\sigma_{in}$ at all four logic points and make the largest of these ratios, $\sigma_{max}/\sigma_{in}$, as small as possible, definitely under 1, by adjusting the tunable parameters. However, this is not always practical, and for many gate functions is not actually mathematically possible. In fact, for our present gate function we can at best get the largest ratio significantly closer to 1 (but still > 1) than we had for the original experimental parameter values. This will become apparent from the derivation given below.

To simplify the calculations, we utilized some specific properties of our gate function. In particular, for both $\alpha$ and $\beta$ large enough, the expression for $F$ can be shown to be approximately symmetric in $\alpha \leftrightarrow \beta$. Our preliminary numerical studies have indicated that the near-optimal parameters should be close to this symmetric situation. This is indeed suggestive because making $\alpha \gg \beta$ or $\alpha \ll \beta$ would unbalance the gradients of $F(x,y)$ near points 01 and 10, making one very small at the expense of the other becoming very large. In fact, our original experimental conditions are in the regime $\alpha \gg \beta$, as can be seen in Figure 4. Thus, we can seek optimal conditions with the assumption that $\alpha = \beta$, which reduces the number of tunable parameters from 3 to 2.

In Figure 6, we plot

$$\Delta\sigma = \left| \left(\sigma_{out}\right)_{at\ 01\ or\ 10} - \left(\sigma_{out}\right)_{at\ 11} \right|, \tag{22}$$

as a function of $\alpha$ and $\tau$. The inset shows the variation of $\sigma_{max}/\sigma_{in}$ in the $\alpha\tau$-plane. The two quantities attain minimum along the same curve, $\tau(\alpha)$, which corresponds to the condition $\left(\sigma_{out}\right)_{at\ 01\ or\ 10} = \left(\sigma_{out}\right)_{at\ 11}$. This curve is not sensitive to the values of $\sigma_{in}$, while the value of $\sigma_{max}/\sigma_{in}$ along the curve is approximately $1.18 + 0.29\sigma_{in}^2$ (for small $\sigma_{in}$).

The fact that $\sigma_{max}$ and $\Delta\sigma$ attain minimum along the same line for our specific gate function, equation 15, is not surprising, because the balance of the gradients near the logic points 01 or 10, vs. that of 11, is determined by the overall degree of convexity of the gate-function cross-sections at fixed $x$ (for 10 vs. 11) and fixed $y$ (for 01 vs. 11). At the same time, for $\alpha = \beta \gg 1$, the convexity (curvature) is controlled by the exponents of the type $\exp(-\alpha\tau x)$, in the numerator of equation 15. Therefore the optimization determines the combination $\alpha\tau$, rather than $\alpha$ or $\tau$ separately, leading to the relation of the form $\alpha\tau = const$. By numerical optimization, we found

$$\alpha = \beta = 1.94/\tau,\tag{23}$$

with high numerical accuracy, see Figure 6. This result determines the optimal values of the biocatalytic reaction rate constants ($R_2, R_3$) vs. the reaction time. The oxygen intake rate ($R_1$) actually cancels out in the product $\alpha\tau$. However, varying it may be helpful in keeping the process parameters within the range of validity of simplified rate-equation approximations.

Let us now discuss the quality of the resulting optimized gate function. It is interesting to compare the probability distributions of input and output for the optimized and our experimentally realized parameter sets. The output distribution can be calculated form the input distributions via the relation

$$Z(z) = \int_0^\infty dx \left[ \frac{X(x)Y(y)}{|\nabla F(x,y)|} \sqrt{1 + \left(\frac{\partial F}{\partial x} \bigg/ \frac{\partial F}{\partial y}\right)^2} \right]_{y \to y(x)},\tag{24}$$

with $y = y(x)$ defined by the fixed-$z$ cross-section, see Figure 5: $z = F(x,y)$. The distributions corresponding to the experimentally realized and optimized gates are illustrated in Figure 7 for inputs 01 and 11. For the "worst-case scenario" input 01, the original gate has a large amplification by the factor up to 4.84. For the optimized gate the noise amplification is significantly reduced, approximately four-fold as compared to the original gate. However, it is still amplification, at both inputs 01, 11, and also at input 10 (not shown in the figure), by about 18% per gate function.

**4. Conclusions and Discussion**

In this work, our experimental data for a biochemical gate was modeled within the rate-equation approach. The resulting rate equations were solved, and the solution was cast in the language of Boolean logic variables. Generally, one can use other phenomenological approaches to map out the "input-output surface." Next, the Boolean inputs and output were treated as analog signals in the context of gate-function optimization. We illustrated a procedure for determining an optimal set for the process parameters, to minimize "analog" noise amplification.

We point out that our original experimental conditions turned out to be quite far from optimal. Our values for the parameter combinations that have to be approximately equal under optimal conditions, namely $\alpha$, $1.94/\tau$, $\beta$, were actually 746, 169, 1.79, respectively. Furthermore, even under the optimal conditions, the present gate function would somewhat amplify "analog" noise. Therefore, a posteriori we have considered it not very useful or illuminating to try to realize the optimal conditions for this particular gate experimentally, especially given the large changes in the parameters required. Instead, we prefer to address, below, criteria and avenues for future work that could lead to more promising biochemical grate realizations that would allow large-scale network realizations.

Specifically, let us assume that a biochemical "circuit" operates in an environment with 5% noise levels, and that we define the tolerance level for the final output (to unambiguously identify 0 or 1) by excluding the middle 1/3 of the interval [0,1]. Then, just two concatenated processing steps with the noise amplification of our original gate will produce too noisy an output to be useful. However, for the optimized conditions we can concatenate gates for up to $p$ processing steps, with the estimate $5\% \times (1.18)^p = 33\%$ yielding approximately $p = 11$. Thus, one can carry out up to order 10 processing steps which is presently safely over the number of concatenated gates in experimentally realized situations.[29,30]

Longer-term, however, the problem of "analog" noise reduction in biochemical logic should be considered within a broader context: Our present study has identified the source of the difficulty as follows. Gate optimization involves modification of the gate function, $z = F(x,y)$, which changes the slope of various cross-sections of this function, some of which were marked in Figure 4(B). Consider the logical input point 01, for instance. In its vicinity, one of the chemicals, $B$, is abundant, but another, $A$, is present in small concentrations (near zero). Thus, the output intensity is simply linear in $A$, along cross-sections that originate near this point. As sketched in Figure 8(A), for larger concentrations of $A$ in our case the cross-section function flattens out (convex function) because the activity of $A$ is decreased due to the required cooperative effect with the "$B$" part of the gate function. Optimization decreases the slope at $A = 0$ but at the expense of having a larger slope at $A = 1$. If instead $A$ has an "autocatalytic" (self-promoter) property (concave function), illustrated in Figure 8(B), then we would be faced with a similar balancing problem.

The standard "way out" of the difficulty is to seek gates with the sigmoid shape of *all* relevant cross-sections. Such a curve, with its curvature (the second derivative) changing sign from positive to negative, is shown in Figure 8(C). In fact, in Figure 4(B) the middle cross-section curve marked, is actually sigmoid (which resulted in the input near 00 not being involved the "gradient balancing" in our optimization), but the two side cross-sections are convex. In the context of biochemical logic the simplest way to advance from a convex case to a sigmoid could be to have some additional process that, instead of decreasing the activity of *A*, simply consumes (deactivates) some of the chemical *A*, but at a rather limited rate or mostly at low concentrations of *A*. Similar to mechanisms recently identified in connection with protein functions in genetic circuits,[41] such processes could compete with the "logic-function" processes and potentially yield a sigmoid response. A similar "removal from the game" should also be designed for the second input, *B*. Of course, this approach was not yet tested, and other mechanisms could be more appropriate.

Finally, in the future it is desirable to focus on those gates that were already realized as a part of concatenated-gate biochemical logic systems. It is best to work with gates that allow for a simple theoretical model, applicable for a wide range of parameters. We expect that candidate gates will function with enzymes and/or enzyme-derived biocatalysts as the "machinery" rather than as inputs. Indeed, simpler chemicals as inputs (and outputs) might offer more flexibility in addressing the aforementioned criteria.

**Acknowledgements.** We gratefully acknowledge support of our research programs by the National Science Foundation under grants CCF-0726698 and DMR-0706209. GS acknowledges Wallace H. Coulter scholarship from Clarkson University. VP wishes to thank Center for Theoretical Biological Physics at University of California, San Diego, for hospitality during the Summer 2007 Workshop "Biological Dynamics of Cellular Processes."

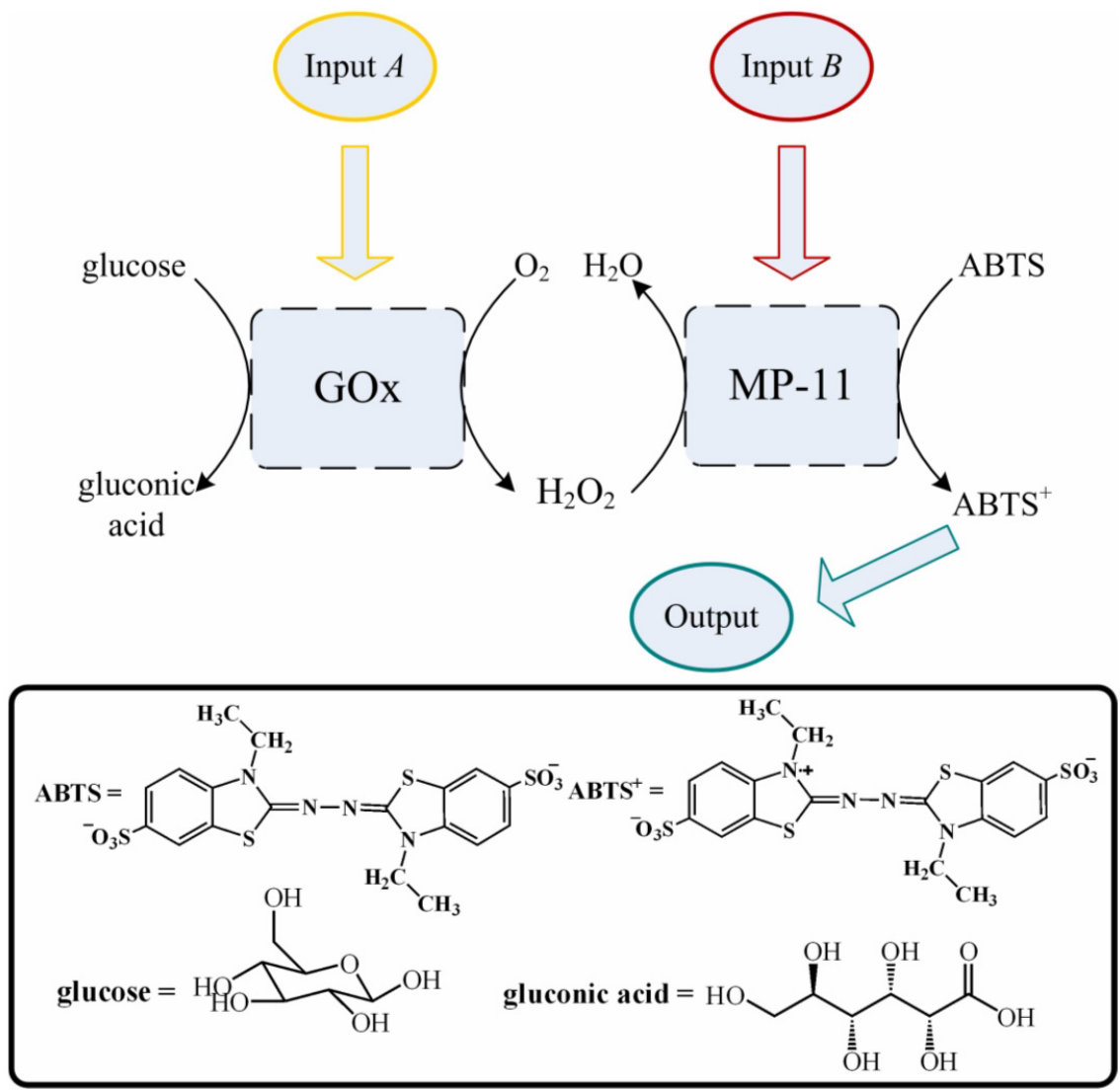

**Figure 1**. Schematic representation of the AND logic gate operating with the two soluble biocatalysts, GOx and MP-11, as the input signals.

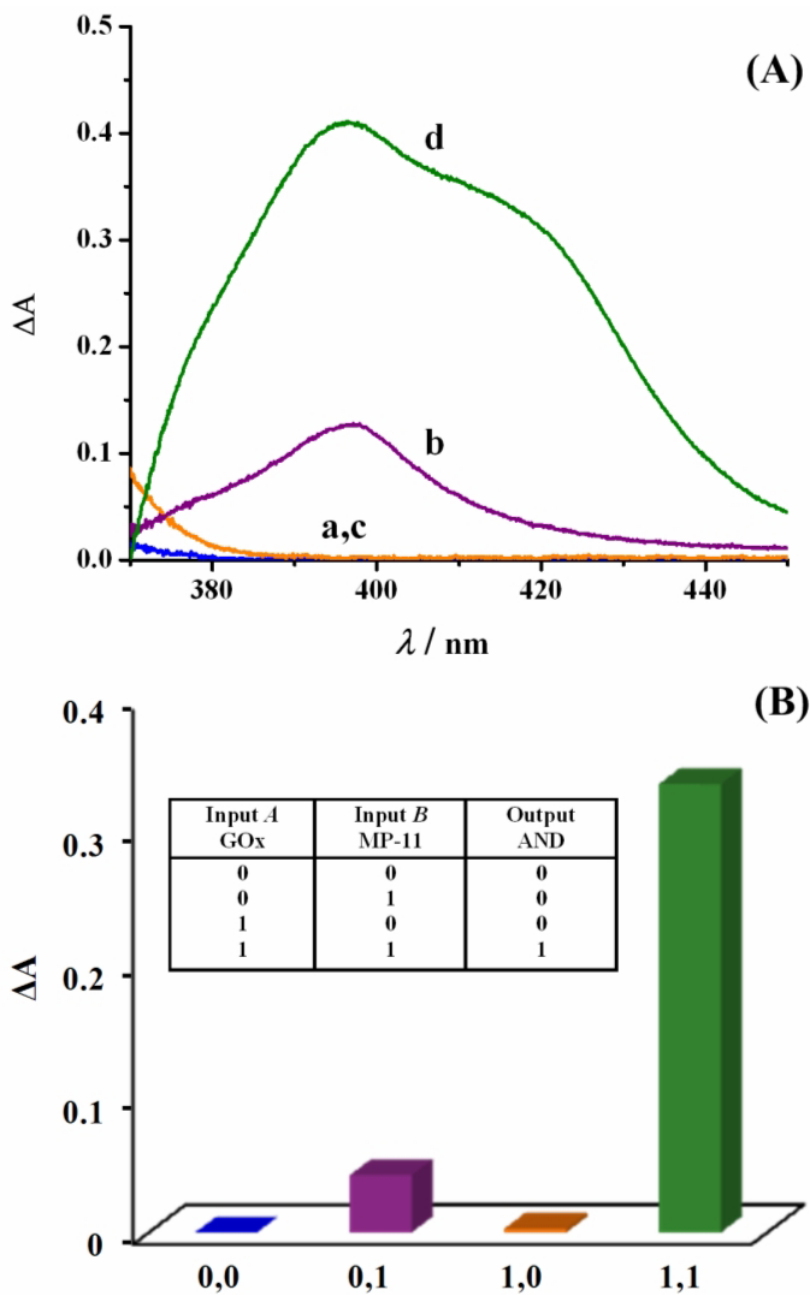

**Figure 2.** (A) Spectral features of the AND gate 145 seconds after the input signals: a) "0,0" – without additions of GOx and MP-11; b) "0,1" – after the addition of MP-11, $1.08\times10^{-6}$ M; c) "1,0" – after the addition of GOx, $1.24\times10^{-8}$ M, d) "1,1" – after the addition of GOx, $1.24\times10^{-8}$ M and MP-11, $1.08\times10^{-6}$ M. (B) Bar chart of the absorbance-deviation outputs at $\lambda = 415$ nm. The inset shows the "truth table" for the AND gate.

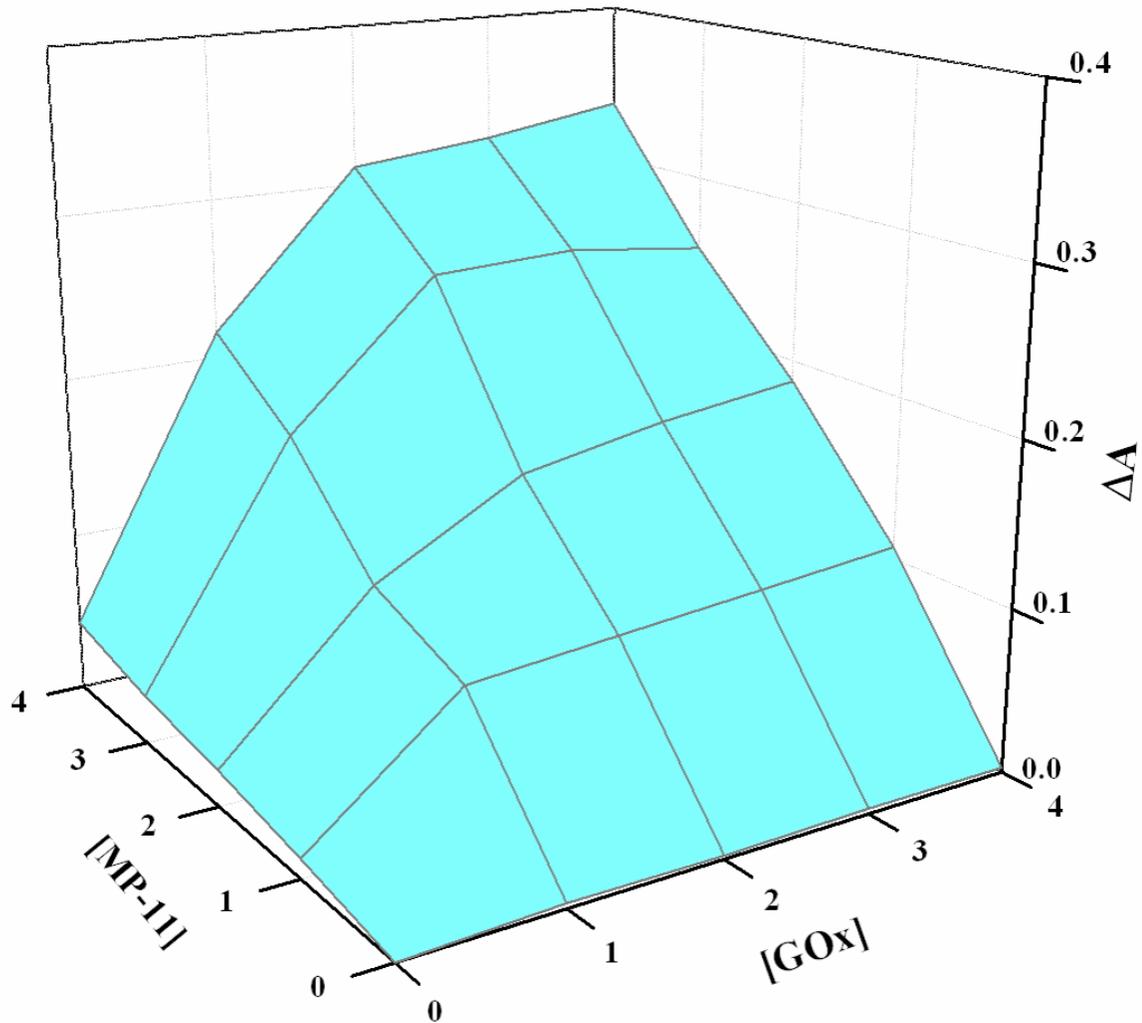

**Figure 3.** Experimental data points shown for various input values studied: The biocatalysts, MP-11 and GOx, when not present in the system corresponded to logic-0 concentrations. Inputs equal to logic-1 were studied for concentrations n = 1, 2, 3 or 4 times $3.1 \times 10^{-9}$ M for GOx, and $2.7 \times 10^{-7}$ M for MP-11. The input axes are labeled by the respective n-values. The different concentrations were combined and then measured for absorbance-deviation output at $\lambda = 415$ nm.

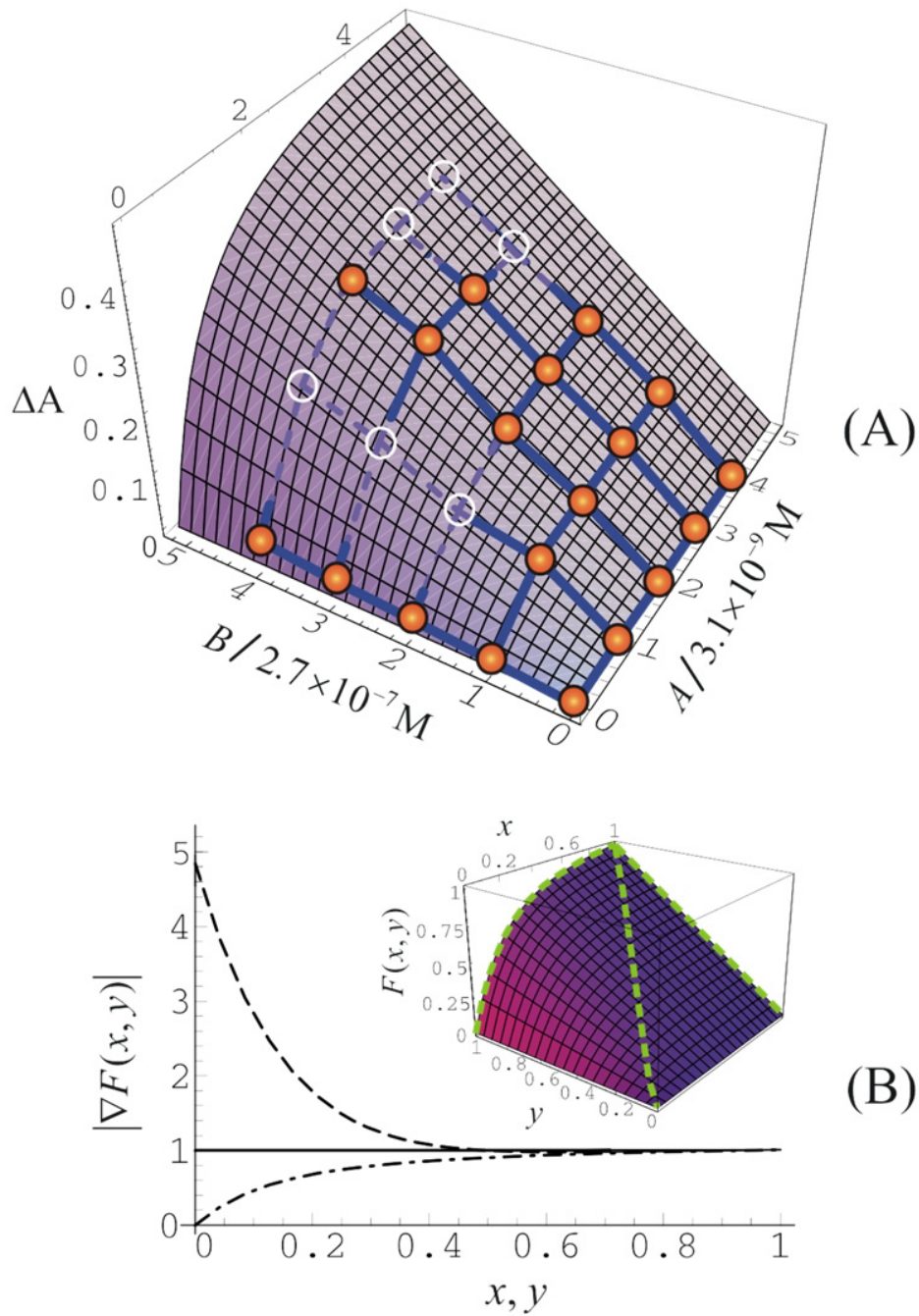

**Figure 4.** (A) The output signal as a function of input concentrations: our theoretical fit to the experimental data points. The experimental points are shown by empty circles when below the function surface. The RMS deviation for the fit is 0.017. (B) Absolute value of the gradient along three cross-sections: 01-11 (dashed curve), 00-11 (dash-doted curve), 01-11 (solid line). The inset shows the gate function in terms of the logic variables, with the three cross-sections shown schematically by the green dashed lines.

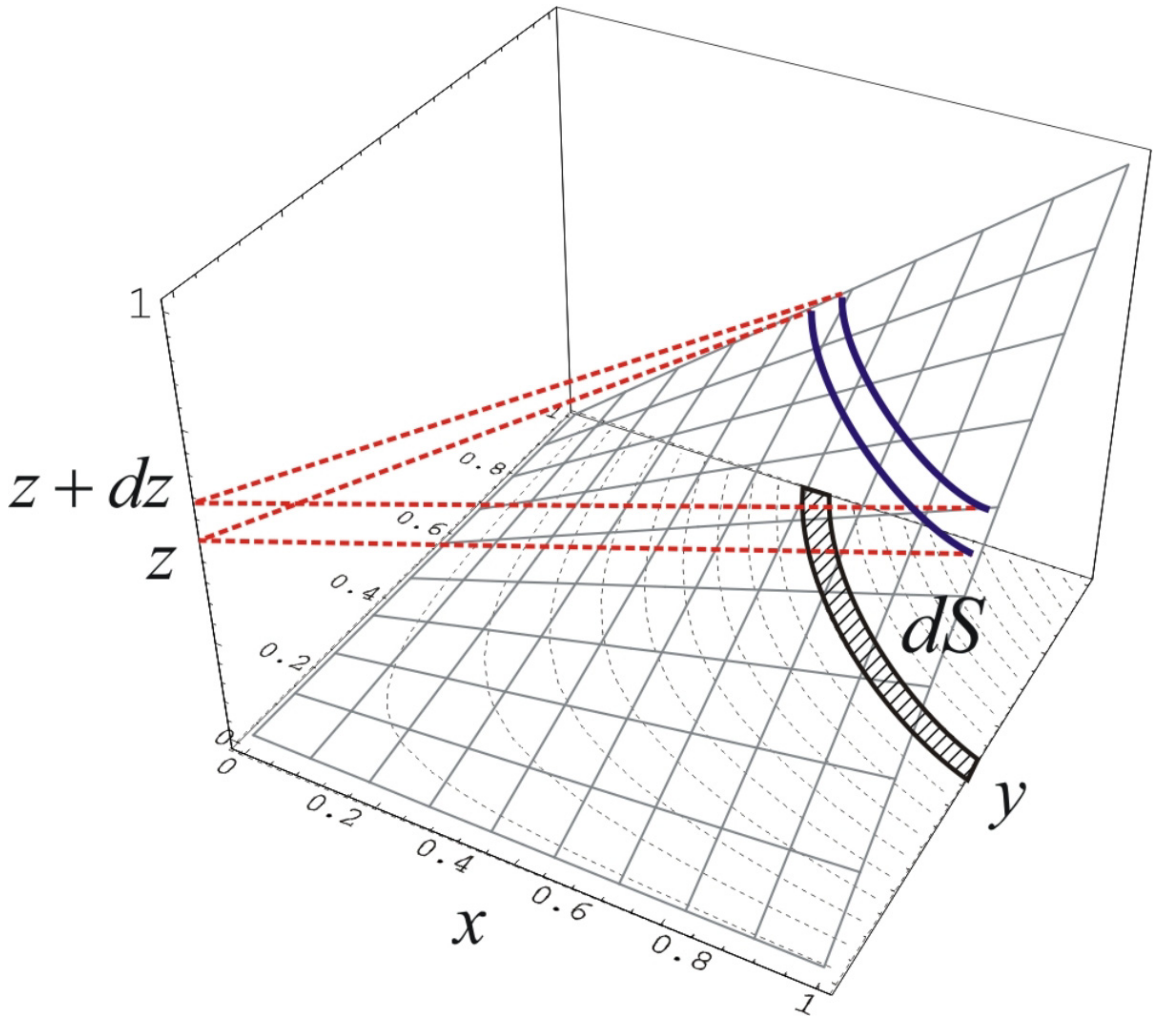

**Figure 5**.  Schematic of the conservation of probability for a gate function: the constant-*z* levels are shown in the *xy*-plane as a contour plot.

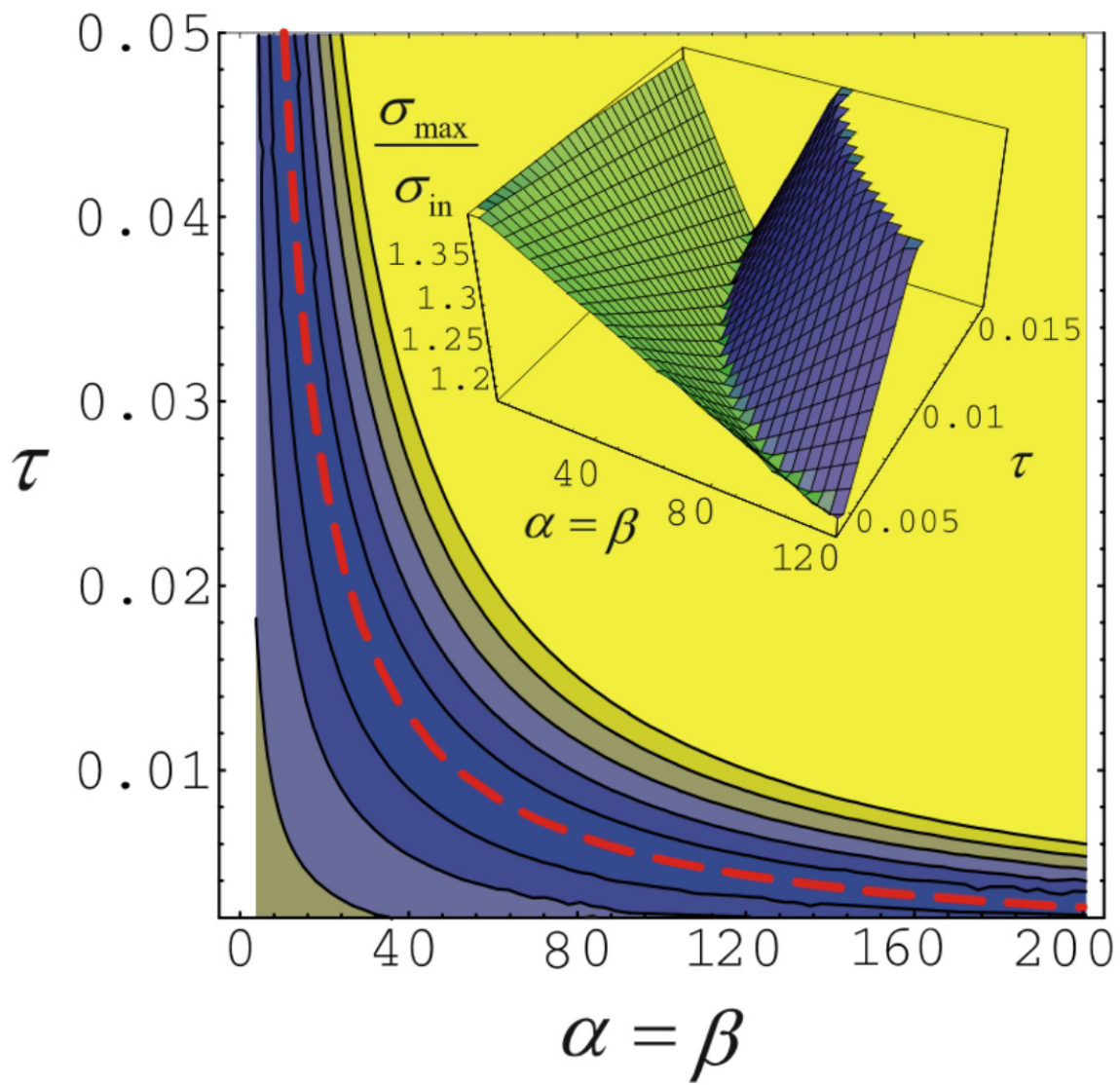

**Figure 6.** Contour plot of the value of $\Delta\sigma$ as a function of $\alpha$ ($=\beta$) and $\tau$. The dashed red line shows the optimum parameters, corresponding to $\Delta\sigma = 0$, which also corresponds to the minimum of $\sigma_{max}$, see the inset. Here $\sigma_{in} = 0.1$, however the optimization line ($\Delta\sigma = 0$) is not sensitive to the value of the input dispersion and has the form $\alpha = \beta = 1.94/\tau$.

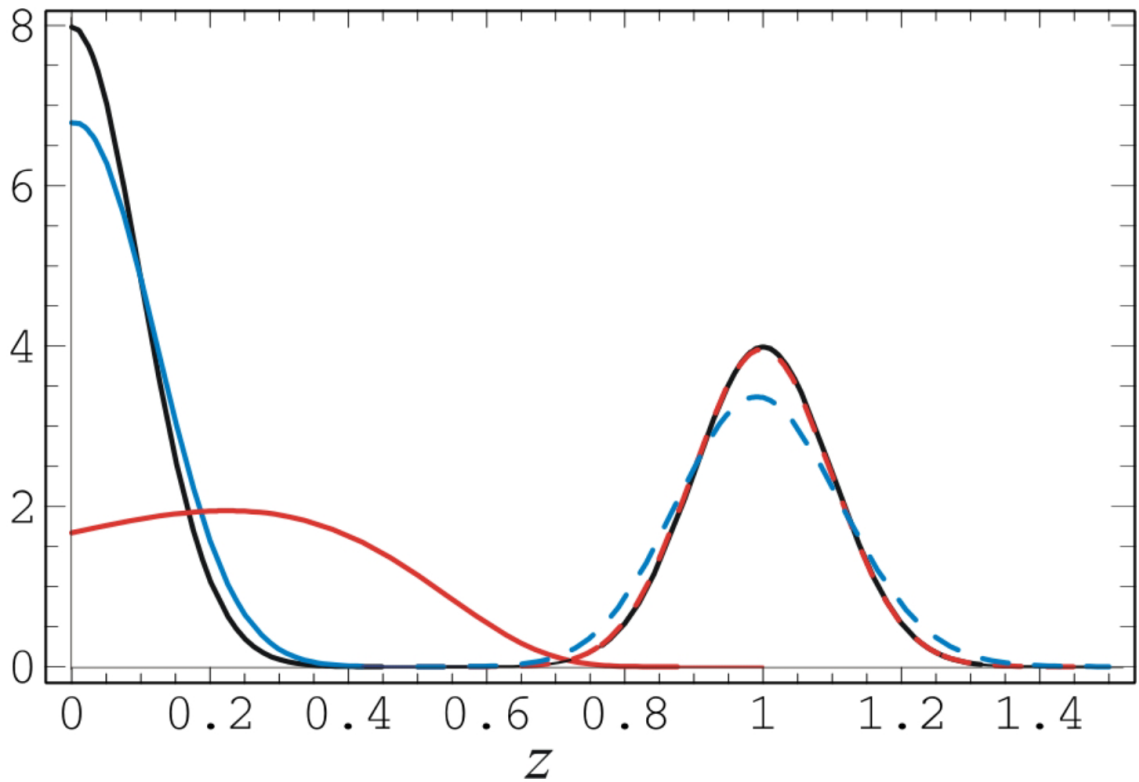

**Figure 7**. Input and output probability distributions for the AND-gate logic action $01 \to 0$ (solid curves) and $11 \to 1$ (dashed curves). The half-Gaussian probability distribution near the input value 0 and the Gaussian distribution near the input value 1, are shown — black curves; both have $\sigma_{in} = 0.1$. For the experimentally realized gate — red curves — the resulting one-sided distribution near the output 0 (for $01 \to 0$) is very broad. Even for the optimized gate — blue curves — both output distributions are also somewhat broadened as compared to the input distributions.

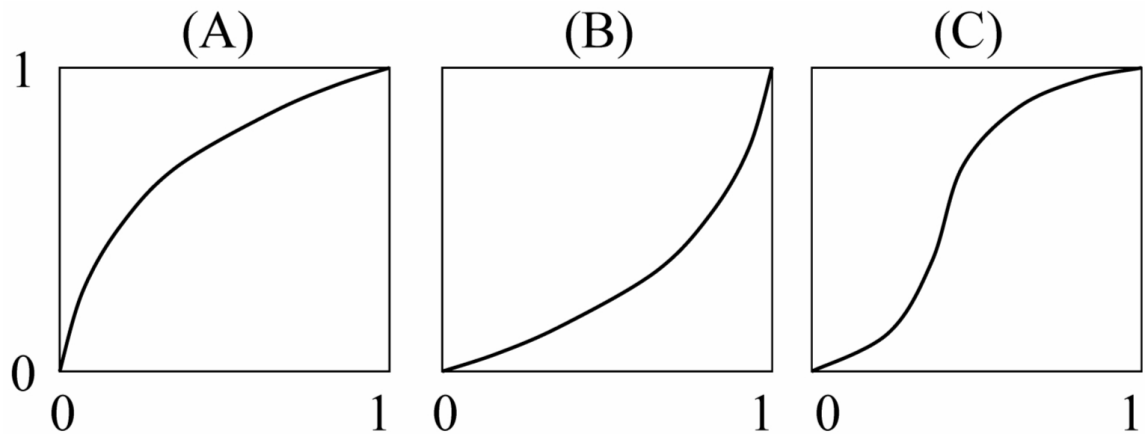

**Figure 8.** Illustration of the (A) convex; (B) concave; and (C) sigmoid shapes of cross-section curves on the "input-output surface," of the type encountered in Figure 4(B).